# Informational interpretation of quantum mechanics

Raoul Nakhmanson

*Nakhmanson@t-online.de*

Ascribing to inanimate matter a possibility to receive, work on and transfer information allows us to explain quantum-mechanical phenomena including "delayed-choice"- and "Einstein-Podolsky-Rosen (EPR)"-type experiments adhering to the basis of local realism, and to suggest essentially new experiments with microparticles and atoms in which information plays the principal rôle.

## I. Introduction

The microworld has surprised the "classical" physicists with the following paradoxes:

• Before quantum mechanics (QM) was created: quantization of mass, charge, energy, angular momentum; the identity of particles of the same type, wave-particle duality, quantum randomness.

• In QM: statistical predictions, Heisenberg's uncertainty principle, Pauli's exclusion principle, and (implicitly but immanently) collapse of the wave function.

• In standard (Copenhagen) interpretation of QM: rejection of the classical realism, a ban on speaking about non-measured parameters, trajectories, etc.; Bohr's complementarity principle.

• In hidden-parameter interpretation: Bell's theorem and non-local instantaneous interaction between far separated microobjects.

• In experiments: results in delayed-choice, EPR, Aharonov-Bohm arrangements.

The Copenhagen interpretation is only a translation of the mathematical formalism of QM to the ordinary language but not an interpretation in a common sense, because it does not explain how, why, and in which framework this formalism works. The common sense of Planck, Einstein, de Broglie, Schrödinger and many others could not accept it: *"ceterum censeo copenhaginem esse delendam"* . But all attempts to find a realistic interpretation were unproductive: they did not forecast new results and introduce doubtful peculiarities e.g. non-local interaction or multiplicating worlds. Feynman told his students that the quantum world was not like anything that we know; and although everybody knows QM, many people use it, some of them develop it, but nobody understands it. Almost equal is Gell-Mann's sentence: "QM, that mysterious, confusing discipline, which none of us really understands but which we know how to use."

How must the productive common-sense interpretation of the quantum mechanics look like?

- It must accept the existence of real and causal world of local-interacting objects;

- It must make quantum mechanics *evident* [1] and explain the above mentioned paradoxes as well as all *"experimenta crucis"* performed up to now;

- It must be verifiable by means of experiments;

- It must show the frame of quantum mechanics and new horizons beyond it.

A version of such interpretation has been developed by the author since 1977 [2-4] and will be summarized as below. It is based on an assumption what microparticles can receive, work on and transfer information, that means, they have some consciousness (spirit, ghost, intelligence), or they are automatons created and programmed from some outside intelligence. This assumption is not a totally new one: just after quantum randomness was discovered people spoke about the "free will" of electrons. Later and more extensively this assumption was discussed e.g. in [5]. But the forerunners did not develop it to the level of a constructive hypothesis that could be tested experimentally.

In Section II the indications for such an assumption are named, the informational interpretation of quantum mechanics is declared and applied to explain the notions of QM and its "paradoxes". In particular, it will be shown that in the new world picture, Bell's theorem [6] is not valid anymore. Besides this there are some remarks about one old paradox of physics, namely, the second law of thermodynamics. Section III considers the results of *"experimenta crucis"* such as those of "delayed choice" [7, 8] and Einstein- Podolsky-Rosen ("EPR") [9, 10] ones. Section IV presents schemes of "Gedankenexperimente" coming out of the frame of QM to have informational contacts with "inanimate" matter. The last schemes do not take into account the direct interaction between matter and human consciousness [11-17] which will be briefly touched in Section V.

## II. Foundation of hypothesis

The idea of microparticle consciousness cannot avoid the following questions:

What clues do we have on the existence of microparticle consciousness?

Have the particles sufficient complexity for it?

What conduced the evolution of microparticle consciousness?

The clues supporting the assumption that "inanimate" matter has some kind of consciousness are the following:

- the formula $E=mc^2$ connecting mass $m$ and energy $E$ by virtue of the light velocity in vacuum $c$ - the maximal velocity of the spreading of information;

- the informational character of the wave function $\psi$ describing probability in contrast to classical potentials describing "tangible" fields transporting energy and impulse;

- "teleological" movement of matter seemingly in the principle of the least action;

♦ quantum-mechanical stochastics which can be seen as optimal tactics of behavior to search for all possible alternatives.

The acceptance of consciousness of microparticles assumes their complex structure. Theoretically it was shown (first perhaps by Markov [18] ) that one microparticle can contain a whole universe ("fridmon", "baby universe"). If one does not go deeper than Planck's length ($\approx 10^{-35}$ m), one finds in a microobject ($\varnothing \approx 10^{-15}$ to $10^{-20}$ m) of the order of $10^{45}$ - $10^{60}$ "Planck cells", that is much more not only than the number of neurons in the human brain, but than the number of atoms contained in all known living beings too. Any detailed assumptions about structures and processes providing the work of particle consciousness are beyond the scope of this article.

The basic circumstance which supports the development of any consciousness is free will. Without free will the consciousness is useless. If free will exists, humans (and not only they) have the choice of alternatives, taking into account physical and social conditions. The more intelligent a choice is, the higher the person's rating in evolution is. Therefore: free will evolves intelligence.

The roots of free will do not lie in the macroworld which is ruled by deterministic laws. They lie in the microworld, and quantum randomness point to it. We cannot be sure that human consciousness is the only product of free will. It is possible that earlier, free will created some consciousness at the level of its roots, i.e., in the microworld. Because the time (measured not in seconds but in events) flowed there much faster, this consciousness had a longer evolution period. Perhaps the golden age of it is over, and now we have to do only with a "rudimentary" consciousness (so called by Cochran, and Bohm and Hiley [5] ) of automatons following the known rules of QM in the majority of natural and experimental situations being observed up to now.

The new conception accepts reality existing beyond our sense-organs and measuring apparatuses and "exculpates" those who say and/or think like "the atom being in point $A$ emits one photon having energy $E$ in the direction $AB$ at the moment $t$ " without being sure whether all these parameters are measured. Moreover, the questions like "But how does your electron know, falling from the upper level, where it has to stop?" (Rutherfort to Bohr, 1911) are permitted. The QM is only a theory of measurements carried out on the microobjects. For example, the accuracy of the simultaneous measurements of complementary variables is restricted because of an interaction with a measuring apparatus that is reflected in Heisenberg's uncertainty principle. But microparticles know their variables, they remember what happened and tell it to others. To do this, they must have synchronized clocks, measuring rules, and reference points for space and time. In this sense it is possible to speak about special ("absolute") coordinates and time, like Greenwich's ones. If we can communicate with particles (see Section IV), the dream of Einstein and other realists, to know the values of all variables included in a theory, can become true.

The Fermi-particles obey the Pauli's exclusion principle. So they can better search for all possibilities. Such a behavior is typical of scientists: each of them tries to find his own theme. Sometimes human's behavior is like a Bose-particle's one. Phenomena such as fashion in dress or music, and applause or coughing in concert halls, are examples of Bose-condensation. The same person can manifest himself as boson or fermion. For particles this was only possible in a "big bang" time. Are we now at the same beginning stage of evolution?

The evolution of the quantum state pursues some purpose that can be formalized into an integral equation via the least action principle. The scientists are successfully guessing a view of the least action function deeper than its meaning. The consciousness of each elementary particle solves this equation taking into account all information which it has. The solution is its own wave function $\psi$ which reflects the strategy of particle behavior. Two or more particles coming once in interaction can have some group strategy and common non-factorized ("entangled") $\psi$-function controlling their behavior in the following near future.

Where is the $\psi$-function? It is not in the real 3-dimensional space. It is in imaginary configurational space, which, in its turn, is in the imagination (consciousness) of the particle. This explains why it does not affect other particles being in the same 3d-space. When the particle receives new information (it can take place by any interaction with micro- or macroobjects), it corrects its strategy. Thus occurs the collapse of the wave function. It occurs not in the real (infinite) space, but in the consciousness of particle. The consequent time is determined by the rapidity of this consciousness. Therefore, compared with space-time conditions of experiment, collapse is local and instantaneous. Von Neumann [11], London and Bauer [12], and Wigner [13] suggested that human consciousness provokes a collapse of $\psi$-function. This is not so: human consciousness collapses only the human knowledge about the $\psi$-function. The laws of both collapses lie beyond physics. The renovation of the knowledge's because of interactions can go without alteration of mass, energy and entropy: the new state of particle can have the same energy and entropy, but different actual information [3].

Through the development of QM the physicists came to the idea about "holism" of the material world. The new interpretation explains this holism as having an informational origin. The world is entangled by information, its "internet" has existed since the big bang.

The particle can know only information lying in the lower light cone, i.e., belonging to the past. To calculate its own strategy (i.e. $\psi$-function) successfully the consciousness of the particle must have an ability to forecast the circumstances in the future using its knowledge about the past. It is the very natural result of development of every consciousness. Particularly, all human life is based on such ability. If a physicist calculates the tomorrow behavior of an elementary particle in his laboratory, he looks into the future, and only a part of it depends on his free will.

In the new conception the paradox of "Schrödinger's cat" [19] vanishes. The cat is either alive or dead, and its state always can be seen by an observer. The mixed "alive-dead" cat exists only as a phantom in the consciousness of a radioactive atom (if it is still potentially active), in the consciousness of the observer (if he has not seen the dead cat), perhaps in the consciousness of the cat, etc., but not in the real space. All of them cannot forecast when the cat will be killed: it depends on the random generator controlling the atom activity (about randomisation of quantum events see below).

The new conception is a "hidden-parameter" one, but these "parameters" are not mechanical ones. But what about Bell's theorem saying that if QM is valid no local hidden-parameter theory is possible?

The proof of Bell's theorem is based on the next assertion: if $P_a$ is a probability of result $a$ measured on the particle $\boldsymbol{1}$ in the point $\boldsymbol{A}$ having a condition (e.g. angle of analyzer) $\alpha$, and $P_b$ is a probability of result $b$ measured on the particle $\boldsymbol{2}$ in the distant point $\boldsymbol{B}$ having a condition $\beta$, then $\beta$ has no influence on the $P_a$, and vice versa. Here Bell and others saw the indispensable requirement of local realism and "separability". Mathematically it can be written as

$$P_{ab}(\lambda_{1i},\lambda_{2i},\alpha,\beta) = P_a(\lambda_{1i},\alpha){\times}P_b(\lambda_{2i},\beta) \qquad \text{(Bell)} \ , \qquad (1)$$

where $P_{ab}$ is the probability of the join result $ab$, and $\lambda_{1i}$ and $\lambda_{2i}$ are hidden parameters of particles $\boldsymbol{1}$ and $\boldsymbol{2}$ in an arbitrary local-realistic theory. Under the influence of Bell's theorem and the experiments following it and showing, that for entangled particles the condition (1) is no longer valid, some "realists" reject locality. In this case an instantaneous action at a distance is possible, and one can write

$$P_{ab}(\lambda_{1i},\lambda_{2i},\alpha,\beta) = P_a(\lambda_{1i},\alpha,\beta){\times}P_b(\lambda_{2i},\beta,\alpha) \qquad \text{(non-locality)} \ . \qquad (2)$$

In principle such a relation permits a description of any correlation between $a$ and $b$, particularly predicted by QM and observed in experiments. But in the frame of local realism the condition (1) is not indispensable. Instead, one can write

$$P_{ab}(\lambda_{1i},\lambda_{2i},\alpha,\beta) = P_a(\lambda_{1i},\alpha,\beta'){\times}P_b(\lambda_{2i},\beta,\alpha') \qquad \text{(forecast)} \ , \qquad (3)$$

where $\alpha'$ and $\beta'$ are the conditions of measurements in points $\boldsymbol{A}$ and $\boldsymbol{B}$, respectively, as they can be forecast by particles at the moment of their parting. If the forecast is good enough, i.e., $\alpha' \approx \alpha$ and $\beta' \approx \beta$, then (3) practically coincides with (2) and has all its advantages plus locality.

On the issue of "separability": The entangled particles have a common strategy and keep it as long as they can forecast the future at the moment of their parting. The new-coming unlooked-for circumstances allow them gradually to cut off and forget the old partnership.

The wave-particle duality is a mind-body one. In the space there exists only the particle; the $\psi$-wave exists in its consciousness, as well as the reflection of the whole world. If there are many particles, their distribution in accordance with the $\psi$-function, e.g. interference fringes, looks like a product of real wave in real space.

Because of free will the behavior of particles is not strictly determined. In situations allowing alternative outputs the theory gives only a distribution of priorities. Taking this into account the particle makes its choice. The optimal tactics of proportional proving of all possibilities is randomization of this choice. It seems as if each particle has its own random generator. The same seems to be valid for people: our unconsciousness often does not choose the "optimal" way but dices obviously using the priorities. However, our consciousness does not yet make sufficient use of proportional randomization [20].

It is known that Einstein did not accept the fortuity implicated by QM: "God does not play dice." Bohr replied: "It is not the job of scientists to prescribe to God how He should run the world." But the matter in question is not God but particles: *they* play dice. As was said, people also use random choice. Some of them believe that so they transfer the choice of decision to God, i.e., God determines the decision and His rôle is exactly the one that Einstein wanted. On the other hand, Einstein believed in Spinoza's God, "which concerns of nature and give up people themselves". If Einstein thought the microparticles might be intelligent, perhaps he would give up microparticles themselves, so that his God can concentrate Himself on something more important.

Very likely the particles are artificial things like some small space ships. Division into different sorts or species with internal identities is typical for mass products. It simplifies production, usage, repairs, and replacement of such objects. Technical objects, plants, animals and humans illustrate this very well. In the last three cases the production is ruled on the genetic level. For example, people have a very narrow statistical distribution of sizes, masses, and performance; the world records in sport differ from the average results not more than twice. The identity of particles of one sort in QM is analogous to the identity of vehicles of one sort with respect to traffic rules. The individual differences (registration number, color, firma-producer, sex and name of driver, etc.) lie beyond the rules. The individual knowledge of elementary particles lie also beyond QM.

Connection between physics and information was touched in latent form in earlier discussions around the second law of thermodynamics and Maxwell's "demon" yet in 19th century. Later it was manifested, e.g., in [21] . One modification of demon's machine is shown in Fig. 1. There is a chamber having two pistons *1* and *2* and filled with gas. The pressure of gas is estimated by atmospheric one, temperature and the weight *W* sitting on the piston *2* . If last three parameters are constant and the chamber has no leakage, the pressure of gas and the volume *V* occupied by gas are constant too.

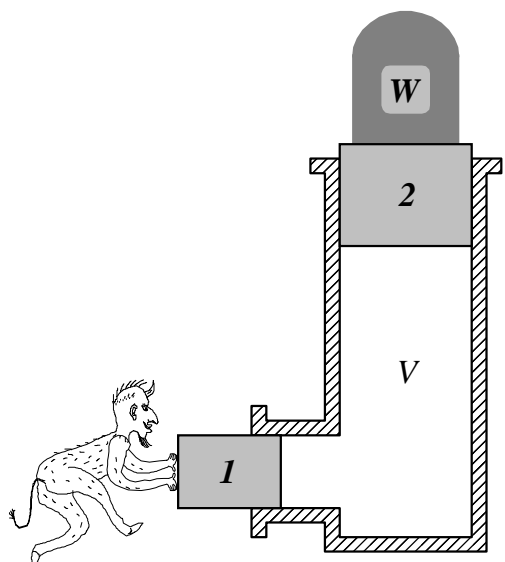

FIG. 1. The chamber with two pistons *1* and *2* contains gas occupying volume *V*. The "demon" moves the piston *1* in time intervals between hits of molecules and lifts the weight *W* by virtue of thermal energy of surrounding only.

The "demon" stays beyond the piston *1* and moves it, but only in time intervals between the hits of molecules. Therefore the demon does not spend any energy to do it.

Because $V$ is constant the piston **2** moves up elevating the weight $W$. The molecules reflecting from moving piston **2** lose their velocity i.e. are cooled, but renew their energy through contact with surrounding. Therefore the demon's machine transform 100% of thermal energy coming from the surroundings into mechanical energy, realizing the "perpetuum mobile of the second kind". If the piston **2** were fixed at the beginning, the gas would be compressed and, without performing any work on the gas, its entropy would be decreased. Both "results" contradict the second law of thermodynamics.

To save the second law of thermodynamics the concept of information was called up [21]. The law was rewritten: instead of $\Delta S \geq 0$ there came

$$\Delta S \geq \Delta I , \qquad (4)$$

where $\Delta S$ is the entropy produced (in our case by demon and his measuring apparatus) to have a new information $\Delta I$ (and to erase the old one). In our last case the demon spends a part of this information $\Delta I'$

$$\left| \Delta I' \right| \leq \left| \Delta I \right| , \qquad (5)$$

to move the piston **1** skillfully to compress the gas and change its entropy $\Delta S'$ without work. Again

$$\Delta S' \geq \Delta I' , \qquad (6)$$

where both partners are negative. Combining (4) to (6) leads to

$$\Delta S \geq \Delta I \geq -\Delta I' \geq -\Delta S' , \ \text{i.e.} \ \Delta S + \Delta S' \geq 0, \qquad (7)$$

i.e. the common entropy increases (or stays constant for reversible processes) in accordance with the (corrected) second law. "We cannot get anything for nothing, not even an observation", - Gabor concluded.

Nevertheless, the second law persists to be a puzzle which cannot be deduced from mechanics with its time-symmetrical laws and, as well as the "time arrow", must be accepted *ad hoc* .

The informational interpretation of QM can open up a new possibility to understand the second law: states with maximum entropy can be preferred by particles for having some purposeful advantages. For example, molecules of gas in a chamber search the space and walls better when they are distributing themselves stochastic-homogeneously, both as an ensemble as well as individuals averaged in time. Accordingly, being injected in the chamber, the molecules try to develop such a distribution faster and keep it forever. In the state with maximum entropy the molecules disperse themselves randomly, but if they discover a deviation from this state they act purposefully and depress the fluctuation using control of the collision parameters (by collisions with walls and with each other) and so producing the macroscopic asymmetry of time.

In the book *The Chance* [22] , issued in 1913, the year of Bohr's atom model, Emil Borel, a notable pure and applied mathematician and an acquaintance of the de Broglie family, remarked that humans cause entropy to decrease in small volumes by means of processes accompanied by increasing of entropy in big volumes, and then went on as follows:

In other words, the structure of the universe is becoming more and more subtle ... , It is probable that similar phenomena are going on other scales as well, too large or too small to be accessible to us. Thus, the evolution of the universe may be represented as a gradual complication of its structure, accessible to understanding and use of beings of lesser and lesser size. As there is no absolute standard of length, we may not be afraid of such a lessening of scales; it seems to us presently that beings of molecular sizes and, all the more, beings so small in respect to molecules as we are to sun, are objects scarcely deserving our attention; but it is quite possible that the progressing complexity of the universe will create or has already created some beings with an organization much more complex than ours.

Following Borel's idea the second law and time arrow in our macroscopical world can also be ascribed to the intellectual activity of microparticles drawing off the "negentropy" in their world (Is part of that drawn off also? [23] ) . Therefore the microworld must own a lot of "negentropy", e.g., information. Because information, more as energy, is the essence of life, can mankind ask microparticles to reverse the second law (and, on occasion, the direction of time) and to return a part of this negentropy to the macroscopical world? To do so, we must have an informational contact with them. Some ideas of such contacts will be presented in Sections IV and V.

### III. *experimenta crucis*

In some experiments performed up to now, the information spreading was implied and one tried to interrupt it, however without success. In others one tried to achieve the information about a material object without having any material contact with it. Below we discuss four of the experiments from the new point of view.

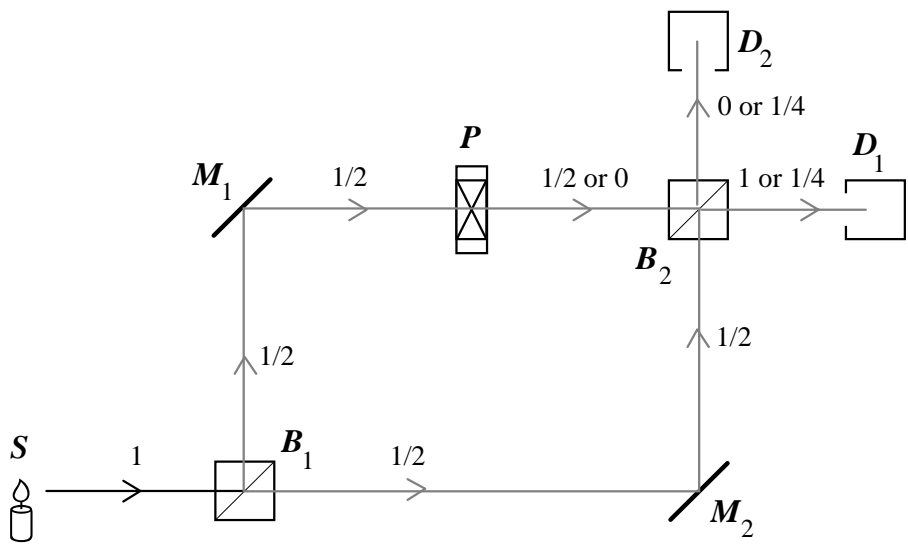

FIG. 2. "Delayed-choice" experiment with Mach-Zender interferometer. *S* is the source of photons, $B_1$ and $B_2$ are the beamsplitters, $M_1$ and $M_2$ are the mirrors, $D_1$ and $D_2$ are the detectors, *P* is the Pockels cell. The ciphers indicate probabilities to find a photon.

• Delayed-choice experiment. It was suggested by Wheeler [7] and really performed in Marylend [8]. Fig. 2 shows the schema. The particles (photons) coming from source *S* enter into the Mach-Zender interferometer with beamsplitters $B_1$ and $B_2$ , mirrors $M_1$ and $M_2$ and detectors $D_1$ and $D_2$ . In one ("upper") arm of the interferometer there is the Pockels cell $P$ , a very fast shutter controlled by a voltage. The interferometer is tuned so that if the Pockels cell is opened all photons are coming to the detector $D_1$ due to constructive interference. The flow of photons is so weak that no more than one photon in the interferometer is a rule.

The Pockels cell is at all times opened (closed), except for a short time interval $\tau$ (several nanoseconds) when it is closed (opened). As a result of experiments having been performed, the interference after $B_2$ appears only if the Pockels cell was opened during the $\tau$ at the appropriate moment when a photon can pass the Pockels cell, independent of its state at other times.

This result cannot be explained on the level of classical local realism thinking that, after $B_1$ , a photon really chooses one arm (upper or lower) of the interferometer: If the photon really traveled through the lower arm (this "must be" in 50% of the cases), it is too far (several meters) from the Pockels cell, and at the above mentioned "appropriate moment" is under no "influence" by the cell (action spreading with velocity $u>c$ is supposed to be impossible). It must be emphasized that independent of the condition $u>c$ the "influence" was not ascribed to some known physical mechanism and thought more as an information-like abstraction.

From a new point of view all this looks different and very natural. When a photon meets the first beamsplitter $B_1$ it brings new information from the source $S$ . Simultaneously it receives from $B_1$ the fresh information about the *past* of the world, particularly about the interferometer and Pockels cell. The physical conditions at $B_1$ induce a 50% choice. But the decision of photons must be at random: It is the optimal tactics for an ensemble of disconnected photons. The consciousness of the photon works to find an optimal strategy, i.e., $\psi$-function. In this case it is a 50% choice of arm after $B_1$ and interference after $B_2$ , if Pockels cell just appears opened. To find it, the photon solves a variation problem (e.g., the wave equation).

Let us suggest that after $B_1$ the photon takes the lower arm. It meets the mirror $M_2$ and acquires some new information about the *past* of the world. In respect to inteferometer, it is the same information, independent of the state of the Pockels cell in the upper arm, because in the "appropriate moment" the photon is nearer to $M_2$ as $P$ or even already has left $M_2$ . Therefore the part of the wave function of photons relating to the interferometer stays the same. Particularly, if at the beginning the Pockels cell appears to be opened, the photon keeps the idea to interfere after $B_2$ .

Finally, the photon meets the beamsplitter $B_2$ and must again choose its path. Simultaneously it receives from $B_2$ the fresher information about the *past* of the world, including an actual history of the upper arm of the interferometer with the state of Pockels cell at the "appropriate moment". This information is brought to $B_2$ with velocity of light, e.g., by thermal photons emitted by Pockels cell itself: at room temperature the characteristic time $\tau$ of several nanoseconds is enough for $B_2$ to receive from $P$ several thousands of thermal photons. Actually, the number of thermal photons emitted by one and arriving at the other macroobject is

$$N \approx 5 \cdot 10^{10} \, S_1 \, S_2 \, T^3 \, \tau \, / \, L^2 \, ,$$

where $S_1$ and $S_2$ are the effective areas of the first and the second macroobjects, respectively, $T$ is the absolute temperature of $S_1$, $\tau$ is the characteristic time, and $L$ is the distance between $S_1$ and $S_2$. Substituting the typical values $S_1 = S_2 = 2$ cm$^2$, $T = 293°$K, $\tau = 10^{-8}$sec, and $L = 10^3$ cm, one finds $N \approx 5000$.

Now the photon has all necessary information to make a decision, namely, to prefer a direction of constructive interference (if the Pockels cell at the "appropriate moment" was opened) or to make a 50% random choice between two possible directions (if the cell at the "appropriate moment" was closed). If the state of Pockels cell in the "appropriate moment" was different than before, a reduction of the strategy (i.e., a reduction of wave function $\psi$ relating to the experiment) takes place in the consciousness of the photon after interaction with the beamsplitter $\boldsymbol{B_2}$.

To have a success in the "delayed-choice" experiment (i.e., to have a result going outside the frame of QM) one can try to cut off the informational contact between the Pockels cell and the beamsplitter $\boldsymbol{B_2}$, e.g. by introducing a deep-cooled filter which is transparent for optical photons being measured but absorbent for thermal photons. In spite of the fact that all the atoms of the filter would know the state of the Pockels cell, if the filter is cooled by liquid helium, no thermal photons come from it to $\boldsymbol{B_2}$ during the characteristic time $\tau$. Further, one can put the Pockels cell together with its voltage source in a separate electromagnetic screen to cut off the radio channel [24]. Besides, to prevent a forecast of a state of the Pockels cell by $\boldsymbol{B_2}$ and other "member" of the experiment, it is better to control the cell using not a regular but "good" random generator.

• In scheme Fig. 2 one can indicate the existence of a non-transparent object (e.g. closed Pockels cell) if the photon is detected by $\boldsymbol{D_2}$, in spite of the fact that the photon has traveled via the lower arm and does not touch the object. The developing of the theme is co-called "interaction-free measurements" [25]. These experiments can be naturally explained within the scope of the informational interpretation of QM: the photon receives the actual information about the object from the beamsplitter $\boldsymbol{B_2}$. Moreover, because the experiments are performed in stationary conditions, the photon knows all about object just when it is born.

• The situation discussed by Einstein, Podolsky, and Rosen [26], and in modern form by Bohm [27] and Bell [6], is more complex. Here two particles flying in the opposite directions have a common non-factorizable wave function, i.e., common correlated strategy. As a consequence, the result of interaction of one particle with some measurement apparatus is strongly correlated with the result of the interaction of another particle with another measurement apparatus, in spite of a large space separating of the two apparatuses. As it was seemed, such a correlation is possible only if the particles are connected by instantaneous non-local interaction. The informational interpretation offers another possibility: each apparatus and/or the particle interacting with it can *forecast* the state of the other apparatus at the "appropriate moment" of measuring with a good probability. It was mentioned in Section II that such a forecast is the very natural result of evolution of microparticle consciousness. In all EPR-experiments made up to now, such forecasts could take place, without assuming a superluminal velocity of interconnection, because the states of the measurement apparatuses have been changed very slowly [9] or periodically [10] . To go outside the scope of QM we can try to perform EPR-experiments with a "good" random control of measurement apparatuses.

It is interesting to note that the authors of cited experimental researches felt the advantage of random control (as it seems more intuitive, because they do not discuss it) and sometimes used it [8]. The peculiarity of a random signal series is the non-predictability of its next term. Therefore, these authors felt a possibility of "inanimate" matter to forecast the future, and have tried to restrict it [8, 10]. The authors of recent work [28] said it clearly: "Selection of an analyzer direction has to be completely unpredictable, which necessitates a physical random number generator. A pseudo-random-number generator cannot be used, since its state at any time is predetermined" (p.5039). As a "physical random number generator" they used a "light-emitting diode illuminating a beam splitter whose outputs are monitored by photomultipliers" (p.5041).

Such a point of view must be commented. It is a right prerequisite that the generator must be unpredictable for particles, but referring to above "physical" generator is not enough: If particles have "consciousness" such a generator can also be a "pseudo" one. Perhaps a good "human" pseudo-random generator is preferable because it belongs to another civilization.

• Aharonov-Bohm effect [29]. In accordance with QM the frequency of wave-function oscillations depends on the energy. If the particle has different energies in different arms of the interferometer, it leads to an additional phase shift and changes the interference pattern. The experiments were performed with an electron interferometer and a magnetic vector potential and justified the predictions of QM. It is of interest that in the experiments the electrons did not cross the magnetic field. From the old classical point of view it looks like non-local action at a distance. The new interpretation explains it very naturally: electrons know the situation from the time they are emitted. Besides, this effect emphasizes a priority of potential against field (in classical physics they enjoy equal rights). From the new point of view it is natural, because potential just contributes to the action function whose minimum as a function of trajectory is wanted. It should be observed that the idea of *forecasting* the conditions on the trajectory is also included in the least action principle. The change from integral form to a differential one does not remove the problem: the mathematical derivative is an abstraction, the physical (i.e. operational) derivative is connected with two distant points, and if the particle is in one of them, it knows only the past conditions in the second point and must extend this into the present and future. To obtain a "non-QM" result one can try to use fast random switching of the magnetic vector potential.

## IV. Informational experiments

If big strangers from other worlds arrived on the Earth and would like to test human intelligence, they would not, of course, do it by dipping men into a bath, throwing them from the tower of Pisa, or bumping them against each other in colliding beams, i.e., the strangers would not just act in this brutal way in which physicists have treated particles, hitherto. Instead of this they would try to build an informational contact with men. Different levels of reaction on the information being proposed would be expected:

0. No noticeable reaction.
1. A reaction showing that information is being received.
2. A reaction showing that information is being deciphered correctly.
3. Sending of reciprocal signals.

Have physicists been searching "inanimate" matter as a possibility to build an informational contact with microparticles? The answer is "Yes". Moreover, if the particles are intelligent, up to third reaction level and complete duplex communication are possible.

Fig. 3 introduces into this field. Fig. 3(a) shows a "black box" which is tested by the linearly polarized light beam. Inside of the box the beam meets a thick transparent glass plate fixed at the Brewster angle so that all photons pass the box. The glass plate manifest itself physically only by space shift $\Delta z$ and time delay $\Delta t$ of output photons. Further there is a movable mirror ("traffic divarication") which is controlled by the experimenter to turn or not to turn the beam. Such a control is a brutal one like a traffic barrier closing one of two branches of the road.

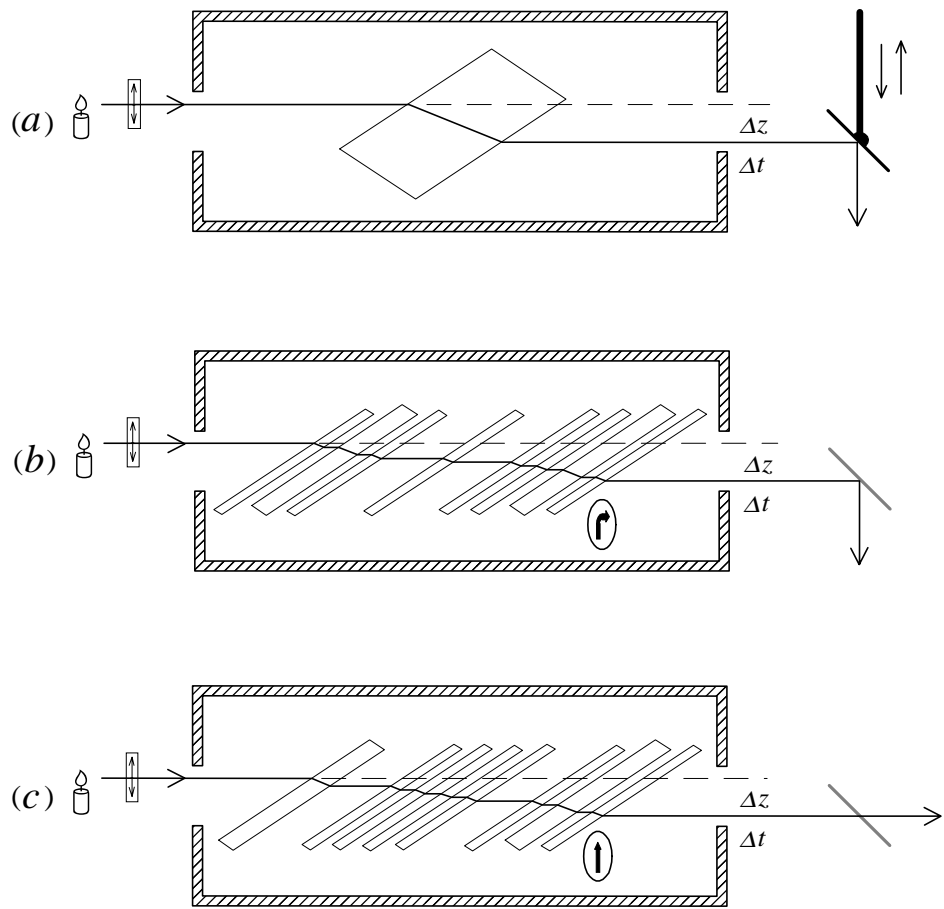

FIG. 3. (*a*) - "brutal" control, (*b*) and (*c*) - informational control .

In Figs. 3(*b*) and 3(*c*) the mirror is semi-transparent and immovable, and the thick glass plate is divided into eight thin plates, two of them being thicker than the remaining six. As before, the content of the black box manifests itself physically only by the same space shift $\Delta z$ and time delay $\Delta t$ . But if the photons are intelligent and know English and Morse code, they can read the messages, namely $\bullet - \bullet \quad \bullet \quad \bullet\bullet - \bullet \quad = \text{REF} \quad (\text{reflect}) \quad \text{in Fig. } 3(b)\,,$

$- \quad \bullet\bullet\bullet\bullet \quad \bullet - \bullet \quad = \text{THR} \quad (\text{through}) \quad \text{in Fig. } 3(c)\,,$

and follow the instructions. Such a control is an informational one like traffic signs on the road.

It is important to emphasize that the idea of informational experiments with particles, as it seems, has never been publicly discussed, and all experiments made with particles up to now cannot be considered as informational ones even in retrospect, that is, the revision of their results would not enable us to make any conclusion relating to this idea.

The way to communication with the particles assumes that they are interested in the information proposed (first level of reaction), that they learn to decipher the information (second level), and that they wish to communicate (third level). The interest in information is thought to be an inherent attribute of each consciousness. One can try to use at the beginning such an attractive "language" as music, and for advance teaching and communication special languages have developed for the project "Search for ExtraTerrestrial Intelligence (SETI)" [30] .

Fig. 4 shows the scheme of "binary-tree" experiment. The initial beam of microobjects (particles, atoms) enters into a system of beamsplitters (shown by circles). They can be semi-transparent mirrors for photons, crystals for electrons or neutrons, Stern-Gerlach apparatuses for atoms, etc. Fig. 4 shows only five rows of beamsplitters, but there can be as many as experimentally feasible. According to present-day theoretical ideas and practical experience, each of the output beams has the same intensity, namely, 1/32 of intensity of the initial beam (real beamsplitters may have, of course, some absorption, but here it is not a matter of principle).

Into each of the right channels of the binary tree is introduced an "informational cell" (shown by birds), which is a device leaving unchanged the intensity of the beam passing it, but offering some information to particles. For polarized photons such a cell may be a set of transparent plates fixed at the Brewster angle, and the information can be coded, say, by differences in the materials of plates, their thickness, and distances between them (see Figs. 3(b), 3(c)) . Absorption caused by the information cells may be compensated by introducing into each of the left channels a "compensating cell" bearing no information, more precisely, bearing less, or less significant information (e.g. the same glass plates placed randomly or periodically). The information in each subsequent row is a sequel of that in the preceding row.

The commonly accepted point of view is that the introduction of information cells, together with compensators, will not change the uniform probability distribution of particles in the output beams. But if particles are intelligent, and are able to notice the information offered to them, they may become interested in it. After a number of rows, the particles should notice that the information is offered only in the right channels, and should prefer the choice of right channels in passing through the following beamsplitters. In other words, particles could develop a "conditioned reflex", of essentially the same kind as in behavior experiments on living beings.

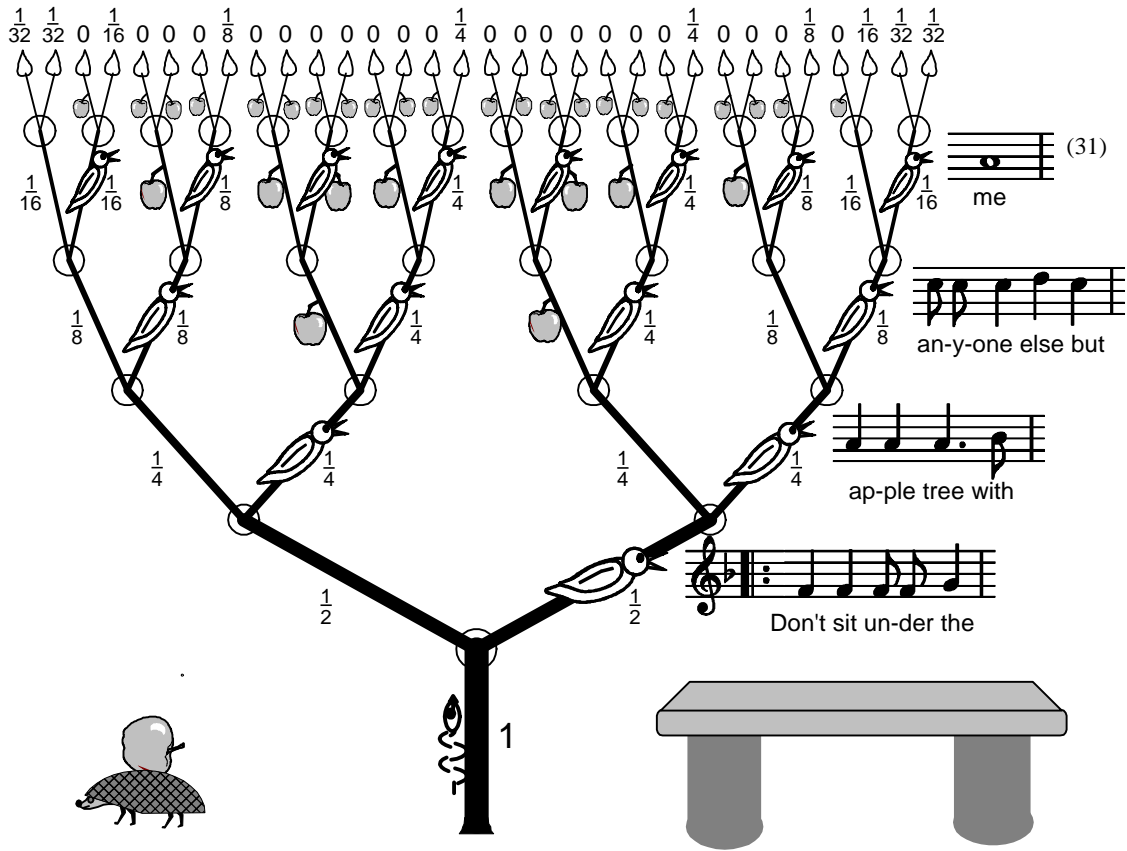

FIG. 4. Binary-tree experiment. Circles stand for beamsplitters, birds denote informational cells, worm denotes the particle. Ciphers and apples (=0) show the probability of detecting the particle in the case of the most rapid formation of a rigid, conservative conditioned reflex. The absorption of the particles by informational cells is assumed to be negligible (i.e. the birds do not eat worms more).

Such an inquisitiveness of particles should lead to a change of their distribution in the output beams. For example, if the conditional reflex appears immediately and the particles are "conservative", i.e., they are no more of interest to the left channels, the distribution of probability to find the particle in different branches of the binary tree is like the one shown in Fig. 4 by ciphers.

Deviation from the uniform distribution of particles in the output beams will mean that the particles at least recognize the information offered and have an interest in it. This, however, still does not mean that the particles understand this information: people of modern times were interested in ancient hieroglyphic symbols long before they learned how to interpret them. To establish a deciphering stage, one can, starting from some row of a binary tree, introduce some specific "requests" into the information cells. For example, one can "ask" particles to choose a left channel after the next beamsplitter rather than a right one. Because between the output branches of the binary tree and the trajectories of the particles there is a one-to-one interrelation, the honoring of such kind of requests can easily be detected by an experimenter. However, the possibilities of an experiment typified in Fig. 4 are not exhausted by this second level of communication. Purposefully choosing direction at each subsequent beamsplitter, the particle, in its turn, can send information to the experimenter using "right" and "left" as a binary code. For example, extreme left and extreme right trajectories in Fig.4 present 00000 = 0 and 11111 = 31, respectively.

The scheme in Fig. 5 shows another possibility. The short light impulse passes through the semi-transparent mirror $E$ and penetrate into a volume limited by two mirrors $A$ and $B$ having reflection near to 100% . If the mirrors are perpendicular to the direction of light propagation, then the light impulse will undergo hundreds of reflections before it is essentially extinguished. At each reflection of the "catched" impulse a small part of the light goes through the mirrors $A$ or $B$ and is registered by detectors $D_a$ or $D_b$ . A semi-transparent mirror $C$ is placed in the middle position between $A$ and $B$ and parallel to them. The impulses registered by detectors follow with time intervals $\Delta t = AB / c$ . Along the optical path between $A$ and $C$ there is an informational cell $I$ controlled from the source of information $SI$ , so that information is renewed with the frequency $1/\Delta t$ . The synchronization is provided from the detector $D_a$ .

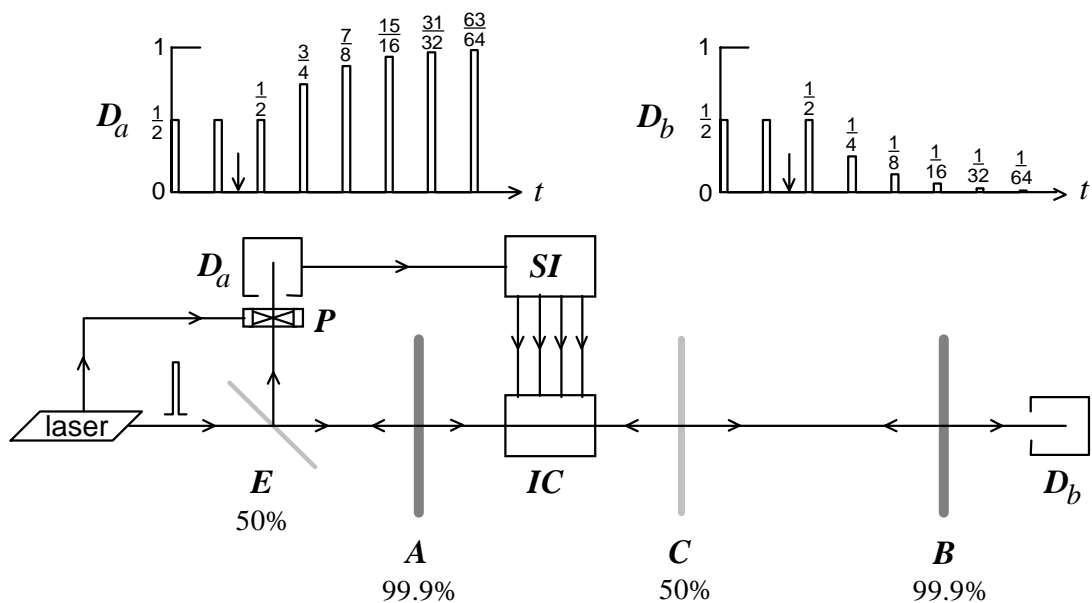

FIG. 5. The "dynamic" informational experiment. $A$ , $B$ , $C$ , and $E$ are mirrors, $D_a$ and $D_b$ are detectors, $IC$ is the informational cell, $SI$ is the source of information. The Pockels cell $P$ is closed only at the short time interval at the beginning to prevent an overload of $D_a$ by the reflection from $A$ . Above, output impulses as functions of time are shown for the case of the first level informational contact and the most rapid formation of a rigid, undisturbable conditioned reflex. The moment the $SI$ is switched on is denoted by arrows. The timing of $SI$ is controlled from $D_a$ .

Before the informational cell $I$ is activated the numbers of photons in the volumes $AC$ and $CB$ are equal and the impulses registered by $D_a$ and $D_b$ have the same amplitudes (the absorption is assumed to be negligible). But after it's activation, and provided the photons are able to perceive the information proposed and to be interested in it, then, after a number of transitions in $AB$ , they should develop a conditioned reflex and, thus, prefer to remain in the volume $AC$ to obtain information. They are able to do so because of their liberty of choice in interacting with the semi-transparent mirror $C$ . This means that the photons would mainly reflect from the mirror $C$ when approaching it from the left, and would mainly go through the mirror $C$ when approaching it from the right. The number of photons in the volume $AC$ and the amplitude of the output impulses beyond the mirror $A$ would increase (twice, at most). The corresponding decrease of the number of photons and of output impulses (to zero as a limit) should be observed for the volume $CB$ . In Fig. 5, at the top, the graphs of intensity of output impulses are represented as functions of time, for the case of informational contact of the first level and of most rapid formation of a rigid, undisturbable conditioned reflex. The moment of activation of the informational cell is shown by arrows.

The scheme Fig. 5 can also be used for informational contacts of second and third levels by redistribution of photons between $AC$ (e.g. representing "0") and $CB$ (representing "1").

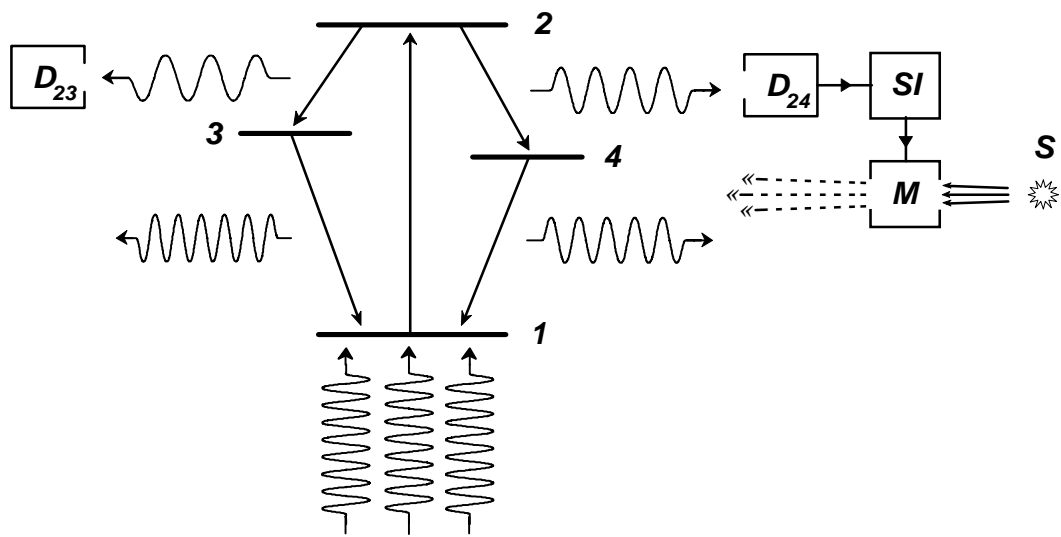

FIG. 6  Informational experiment with a single atom. 1, 2, 3, and 4 are the energy levels; $D_{23}$ and $D_{24}$ are detectors, $S$ is the source of light, $M$ is the modulator, $SI$ is the source of information .

The experiments illustrated in Fig. 4 and Fig. 5 can be called "coordinate-impulse" ones to distinguish them from the "energy-time" experiment whose scheme is shown in Fig. 6 . Here a four-level quantum system, e.g., an atom, with one low (1), one high (2), and two intermediate (3,4) energy levels is pumped by intensive radiation inducing the $1\rightarrow2$ transition, so that the atom stays not in the state 1 but immediately is translated into the state 2. From it, the atom makes transition spontaneously to the state 3 or 4, and later makes transition to the state 1 completing the cycle. The radiation corresponding to some of transitions $2\rightarrow3$, $2\rightarrow4$, $3\rightarrow1$ and $4\rightarrow1$ are detected (in Fig. 6 two detectors are shown). Besides, there is an informational action on the atom, e.g., by modulation of light coming from the source $S$ . The modulator $M$ is controlled by the source of information $SI$ , which, in turn, is connected with one or more detectors to close the feedback loop.

The feedback works in such a way as to stimulate a channel and rate of transitions, in the case of Fig. 6, the $2\rightarrow4\rightarrow1$ transitions. The source $SI$ sends a message, e.g., one line of a page or a measure of a music, only if it receives a signal from detector $D_{24}$ . Each next message continues the previous one, i.e., is the next line or the next measure.

If the atom has an intelligence and is interested in the information being proposed, it develops a conditioned reflex and will prefer the 2→4 transition to the 2→3 one. Besides, the rates of both 2→4 and 4→1 transitions must increase. All this can be registered by the experimenter. To be sure that the effect is connected with information, one can make a control experiment to cut off the feedback or/and to use some "trivial" information, etc.

If the detector $D_{24}$ has a small aperture, the spatial orientation of emitted photons toward to $D_{24}$ may also develop itself as a part of conditioned reflex. In spite of this interesting possibility, one must prefer to use effective detection of emitted photons to facilitate the development of a conditioned reflex. Perhaps the combination of an ion trap and resonator [32] provides a good opportunity for informational experiments with single atoms.

A progress in semiconductor device technology possibly can also be used for such an "energy-time" experiment. In a small metal-oxide-silicon field-effect transistor (MOSFET) one can observe random telegraph signals corresponding to the charge and discharge of a single electron trap located at the $Si-SiO_2$ interface [33]. As in the previous case, the new information may be sent to the MOSFET (i.e. to the trap) by modulation of light or sound only after the next capture or/and emission of an electron by the trap. It closes the feedback loop, and the experimenter may find an increase in the capture and emission rate.

Like with the schemes of Fig. 4 and Fig. 5, in two last cases one may hope to observe not only an interest of a quantum object (atom, electron trap) to receive a new information, but deciphering it also, as well as the sending of messages from the object to the experimenter being coded in states of the atom or electron trap and time intervals between the states.

# V. Informational world of matter
# and human consciousness

If the microparticles and atoms are really "intelligent", why they do not manifest themselves more obviously? For example, why do molecules of air in the room not congregate themselves near radio or TV to hear music and to watch a movie? A semantic answer could be, e.g., that they are not interested in it. Physicists can also take into account the large difference in spatial and temporal scales between structures analogous in their intelligence functions, say, something like the relation between the size of the solar system and that of the atom ($10^{21}$). The molecules of air have millions of collisions per second, and each of them brings a lot of information. In the context the weak and slow alternations of pressure (due to music) or pixel brightness (on TV screen) are noteless. Therefore special conditions and arrangements as described in the previous Section supported by screening and cooling are needed. They can serve not only to have an informational contact with separate particles and atoms. Because it appears that the material world is entangled by information, the above-mentioned arrangements can also serve as interfaces to the "internet of matter".

It is not to exclude that the evolution of plants, animals and men had found possibilities to develop the interfaces connecting them to the ghost of matter. Such a connection would satisfy not only their inquisitiveness but would be useful also. Those who believe in parapsychological phenomena (among them were many notable physicists), as well as scientists searching mind-matter connections [14-17], can find a solid base in the informational interpretation of QM. For example, information must spread in an internet of matter without an attenuation law $1/r^2$ and pass through usual screens - the peculiarities ascribed to these phenomena. Telepathy and clairvoyance can use the internet of matter to send and receive information, and forecasting can use in addition the corresponding presages of particles. The known descriptions of telekinesis implicate the appearance of mechanical energy. Such an energy can be transformed from thermal energy with apparent violation of the second law of thermodynamics (see end of Section II).

It is known that people have several levels of memory. The first one registers all information received by the sense-organs but keeps it only for a short time. After selection the essential part of it is transferred into the second level, etc. The more important the information the deeper level it reaches and the longer it is conserved there. The deepest known level is the genetic one. The idea of informational contact between the world of matter and living beings permits the existence of deeper levels of memory situated in the atoms and elementary particles of these beings [34]. If this is so, the beliefs in souls floating in air and in reincarnation have chances to be accepted by *philosophiae naturalis* .

## VI. Conclusion

In this paper the informational interpretation of quantum mechanics was presented. It is realistic with local interactions, it is in agreement with common sense, it allows experimental verification and explains the quantum paradoxes very naturally. It sheds new light to problems of thermodynamical irreversibility and mind-matter interaction, and extends the field for scientific, technical, science-fiction, and religious speculations. Anyone of these perspectives is interesting enough to prove it.

As it seems not only Boltzmann-Shannon but semantic information also must be introduced in physics. Was it really a Word at the beginning, if it was?

# References

[1]  "... we want more than just a formula. First we have an observation, then we have numbers that we measure, then we have a law which summarizes all the numbers. But the real *glory* of science is that *we can find a way of thinking* such that the law is *evident*."
  *The Feynman lectures on physics*  (Addison-Wesley, MA, 1966), p.26-3.

[2]  R.Nakhmanson, *Preprint 38-79*  (Institute of Semiconductor Physics, Novosibirsk, 1980); see also A.Berezin and R.Nakhmanson. *J. of Physics Essays*  **3**,  331 (1990).

[3]  R.Nakhmanson, in *Waves and Particles in Light and Matter*,  ed. by A.van der Merwe and A.Garuccio  (Plenum Press, New York, 1994), p.571.

[4]  R.Nakhmanson, in *Frontiers of Fundamental Physics*,  ed. by M.Barone and F.Selleri (Plenum Press, New York, 1994), p.591.

[5]  G.Cocconi, in *Evolution of Particle Physics*, ed. by M.Conversy  (New York, 1970); A.A.Cochran, Found. Phys. **1**, 235 (1971);  J.E.Charon, *L`esprit, cet inconnu*  (Albin Michel, Paris, 1977);  *The Ghost in the Atom*, ed. by P.C.W.Davies and J.R.Brown (Cambridge University Press, Cambridge, 1986);  D.Bohm and B.J.Hiley, *The undivided universe*  (Routledge, London, 1993).

[6]  J.S.Bell, Physics **1**, 195 (1964).

[7]  J.A.Wheeler, in *Mathematical Foundations of Quantum Theory*, ed. by A.R.Marlow (Academic Press, New York, 1978), p.9.

[8]  C.O.Alley, O.G.Jakubowicz, and W.C.Wickers, in *Proceedings of 2nd International Symposium on Foundations of Quantum Mechanics*, ed. by M.Namiki *et al.* (Physical Society of Japan, Tokyo, 1987), p.36.

[9]  S.J.Freedman and J.F.Clauser, Phys.Rev.Lett. **28**, 938 (1972).

[10]  A.Aspect, J.Dalibard, and G.Roger, Phys.Rev.Lett. **49**, 1804 (1982).

[11]  J. von Neumann, *Die mathematischen Grundlagen der Quantenmechanik*  (Springer, Berlin, 1932).

[12]  F.London and E.Bauer, *La théorie de l'observation en mécanique quantique*, No.755 of *Actualités scientifiques et industrielles: Exposés de physique générale*  (Hermann et Cie, Paris, 1939),  engl. transl. in *Quantum Theory and Measurement*, ed. by J.A.Wheeler and W.H.Zurek  (Princeton University Press, Princeton, NJ, 1983).

[13]  E.P.Wigner, in *The Scientist Speculates*, ed. by I.J.Good (London, Heinemann, 1961); see also E.P.Wigner, *Symmetries and Reflections*  (M.T.I. Press, Cambridge, MA, 1970).

[14]  L.Bass, Found.Phys. **5**, 159 (1975).

[15]  J.Hall, C.Kim, B.McElroy, and A.Shimony, Found. Phys. **7**, 759 (1977).

[16]  R.G.Jahn and B.J.Dunne, Found. Phys. **16**, 721 (1986).

[17]  D.I.Radin and R.D.Nelson, Found. Phys. **19**, 1499 (1989).

[18]  M.A.Markov, *On the Nature of Matter*,  (Moscow, 1963, in Russian);  id., Ann. Phys. **59**, 109 (1970).

[19]  E.Schrödinger, Naturwiss. **23**, 807 (1935).

[20]   The roulette, lottery, Monte-Carlo method of numerical calculation etc. are without question.  The election of Doge in Venice (697-1797 a.C.) was going in ten steps, five of them used random event generators.  This experience was successful: only one of doge acted *ultra vires* .  If the democracy means the same chance for everyone to realize his program, the election must serve only to estimate the priorities ($\psi$-function), not the decision, otherwise we have the dictatorship of majority.  The decision must be estimated by proportional random procedure.  Thereto such a procedure makes coalitions and wars for 51% meaningless.  If Buridan's ass ventured to use a random procedure in choosing his bundle of hay, he would thus "quantize" himself into one of the two states, and thus escape starvation (noted by V.G.Yerkov).

[21]  L.Szillard,  Zs. f. Physik **53**, 840 (1929);  D.Gabor, *MIT Lectures*  (Massachusetts, 1951); L.Brillouin, *Science and Information Theory*  (Academic Press, New York, 1970); I.Prigogine and I.Stengers, *Order out of Chaos*  (Shambhala, Boulder, 1984).  It must be noted, that the matter of discussion cited is information defined by Shannon's formula (which coincides, within the sign, with Boltzman's one for entropy), not the semantics.

[22]  E.Borel, *Le Hasard*  (Paris, 1913).

[23]  "Great fleas have little fleas,
           Upon their backs to bite  'em;
           Little fleas have lesser fleas,
           And so *ad infinitum* ."
                      ( J. Swift )

[24]  The efficiency of other fast channels (neutrinos, gravitons, ...?) can not be excluded.

[25]  A.C.Elitzur and L.Vaidman, Found. Phys. **23**, 987 (1993);   P.Kwiat, H.Weinfurter, T.Herzog, A.Zeilinger, and M.Kasevich, Ann. NY Acad. Sci. **755**, 383 (1995);  A.G.White, J.R.Mitchell, O.Nairz, and P.G.Kwiat, Phys. Rev. A **58**, 605 (1998).

[26]  A.Einstein, B.Podolsky, and N.Rosen, Phys. Rev. **47**, 777 (1935).

[27]  D.Bohm, *Quantum Theory*  (Prentice Hall, New York, 1951).

[28]   G.Weihs, T.Jennewein, Ch.Simon, H.Weinfurter, and A.Zeilinger, Phys. Rev. Lett. **81**, 5039 (1998).

[29]  Y.Aharonov and D.Bohm, Phys. Rev. **115**, 485 (1959).

[30]  H.Freudental, *LINCOS: Design of a Language for Cosmic Intercourse*  (North-Holland, Amsterdam, 1960).

[31]  L.Brown, C.Tobias, and S.H.Stept, *Don't Sit Under The Apple Tree* .

[32]  G.Rempe, W.Schleich, M.O.Scully, and H.Walther, in *Proceedings of 3rd International Symposium on Foundation of Quantum Mechanics*  (Physical Society of Japan, Tokyo, 1989).

[33]  K.S.Ralls, W.J.Skocpol, L.D.Jackel, R.E.Howard, L.A.Fetter, R.W.Epworth, and D.M.Tennant,  Phys. Rev. Lett. **52**, 228 (1984).

[34]   From this point of view the inquisition had found the best way to spread heresy by burning.